\documentclass[%
 reprint,
 amsmath,amssymb,
 aps,
]{revtex4-1}

\usepackage{graphicx}% Include figure files
\usepackage{dcolumn}% Align table columns on decimal point
\usepackage{bm}% bold math
\usepackage{textcomp}
\usepackage{units}

\begin{document}

\preprint{APS/123-QED}

\title{A compact electron matter wave interferometer for sensor technology}

\author{A. Pooch$^1$, M. Seidling$^1$, M. Layer,  A. Rembold$^1$ and A. Stibor$^{1,+}$}
\affiliation{$^1$ Institute of Physics and Center for Collective Quantum Phenomena in LISA$^+$,
University of T\"{u}bingen, Auf der Morgenstelle 15, 72076 T\"{u}bingen, Germany}

\date{\today}

\begin{abstract}
Remarkable progress can be observed in recent years in the controlled emission, guiding and detection of coherent, free electrons. Those methods were applied in matter wave interferometers leading to high phase sensitivities and  novel sensor technologies for dephasing influences such as mechanical vibrations or electromagnetic frequencies. However, the previous devices have been large laboratory setups. For future sensor applications or tests of the coherence properties of an electron source, small, portable interferometers are required. Here, we demonstrate a compact biprism electron interferometer that can be used for mobile applications. The design was optimized for small dimensions by beam path simulations. The interferometer has a length between the tip and the superposition plane before magnification of only \unit[47]{mm} and provides electron interference pattern with a contrast up to \unit[42.7]{\%}. The detection of two dephasing frequencies at 50 and \unit[150]{Hz} was demonstrated applying second order correlation and Fourier analysis of the interference data.

\end{abstract}

\maketitle

\section{Introduction}

Matter wave interferometers for electrons \cite{Hasselbach2010,Mollenstedt1956a,Hasselbach1988,Schuetz2014} have significantly improved in the last decades. They are applied to measure the rotational phase shift due to the Sagnac effect \cite{Hasselbach1993}, to study Coulomb-induced quantum decoherence \cite{Zurek2003,Sonnentag2007}, the magnetic Aharonov-Bohm effect \cite{Aharonov1959,mollenstedt1962,Tonomura1986,Chambers1960} or the Talbot-Lau effect for magnetic field sensing \cite{Bach2013}. The topic is influenced by recent technical innovations and improvements concerning the beam source \cite{Kuo2006,Ehberger2015,Hommelhoff2006a}, the precise electron guiding \cite{Hasselbach1988,Hammer2014}, the coherent beam path separation \cite{Schuetz2014,Ehberger2015,Chang2009,Cho2004} and the development of spatial and temporal single-particle detection methods \cite{Guenther2015,Rembold2014,Rembold2017,Jagutzki2002}. The progress has potential novel applications in electron microscopy \cite{Putnam2009,Kruit2016} and sensor technology for inertial forces \cite{Clauser1988}, mechanical vibrations \cite{Rembold2017} and electromagnetic frequencies \cite{Guenther2015,Rembold2014}. 

Small deviations of the partial waves in the two separated beam paths in a matter wave interferometer lead to a clear phase shift on the detector after they get superposed. This simple feature makes interferometric measurements extremely sensitive towards external perturbations. In contrast to neutral atoms, the phase of electron matter waves can be shifted not only by mechanical vibrations, temperature drifts or rotations of the setup but also by external electromagnetic frequencies. Usually, these perturbations lead to a time dependent dephasing, causing a ``wash-out'' of the temporally integrated interference pattern that can be observed in a reduced interference contrast. This is particularly a challenge for sensitive long-time phase measurements such as proposed for the measurement of the electric Aharonov-Bohm effect \cite{Schuetz2015b}, for decoherence measurements \cite{Sonnentag2007} or the interferometry of ions \cite{Hasselbach1998a,Maier1997}. 

We recently demonstrated in a biprism electron interferometer \cite{Schuetz2014} that such dephasing effects can on the one hand be corrected and on the other hand used for an accurate measurement of the perturbation frequencies \cite{Guenther2015,Rembold2017}. Thereby, the dephasing was detected and reduced with the high spatial and temporal single-particle resolution of a delay line detector \cite{Jagutzki2002}. A second-order correlation analysis in combination with a Fourier analysis was performed on the detection events after the interference was recorded. It can reveal multifrequency electromagnetic oscillations and mechanical vibrations. The spectrum of the unknown external frequencies, their amplitudes, the interference contrast and the pattern periodicity can be extracted from a spatially ``washed-out'' pattern \cite{Rembold2014,Guenther2015,Rembold2017}. For that reason electron matter wave interferometers have a high potential in sensor technology. However, due to their large dimensions current experimental setups are not suitable for portable sensor applications \cite{Mollenstedt1956a,Hasselbach1988,Schuetz2014}. To apply an electron interferometer as a sensor for electromagnetic and vibrational frequencies or for the mobile analysis of the coherence of a beam source, it is necessary to construct a small and transportable device.

In this article we present a compact biprism matter wave interferometer for free electrons with minimized distances between all parts and a high mechanical stability. The distance between the tip and the superposition plane before magnification is only \unit[47]{mm} including all components for beam guiding and diffraction. In combination with the recently developed tools for spectrum analysis by correlation theory \cite{Rembold2014,Guenther2015,Rembold2017}, the compact setup is an important prerequisite for a portable, mobile sensor based on matter wave interferometry with electrons.

\begin{figure}[t]
\centering
\includegraphics[width=0.45\textwidth]{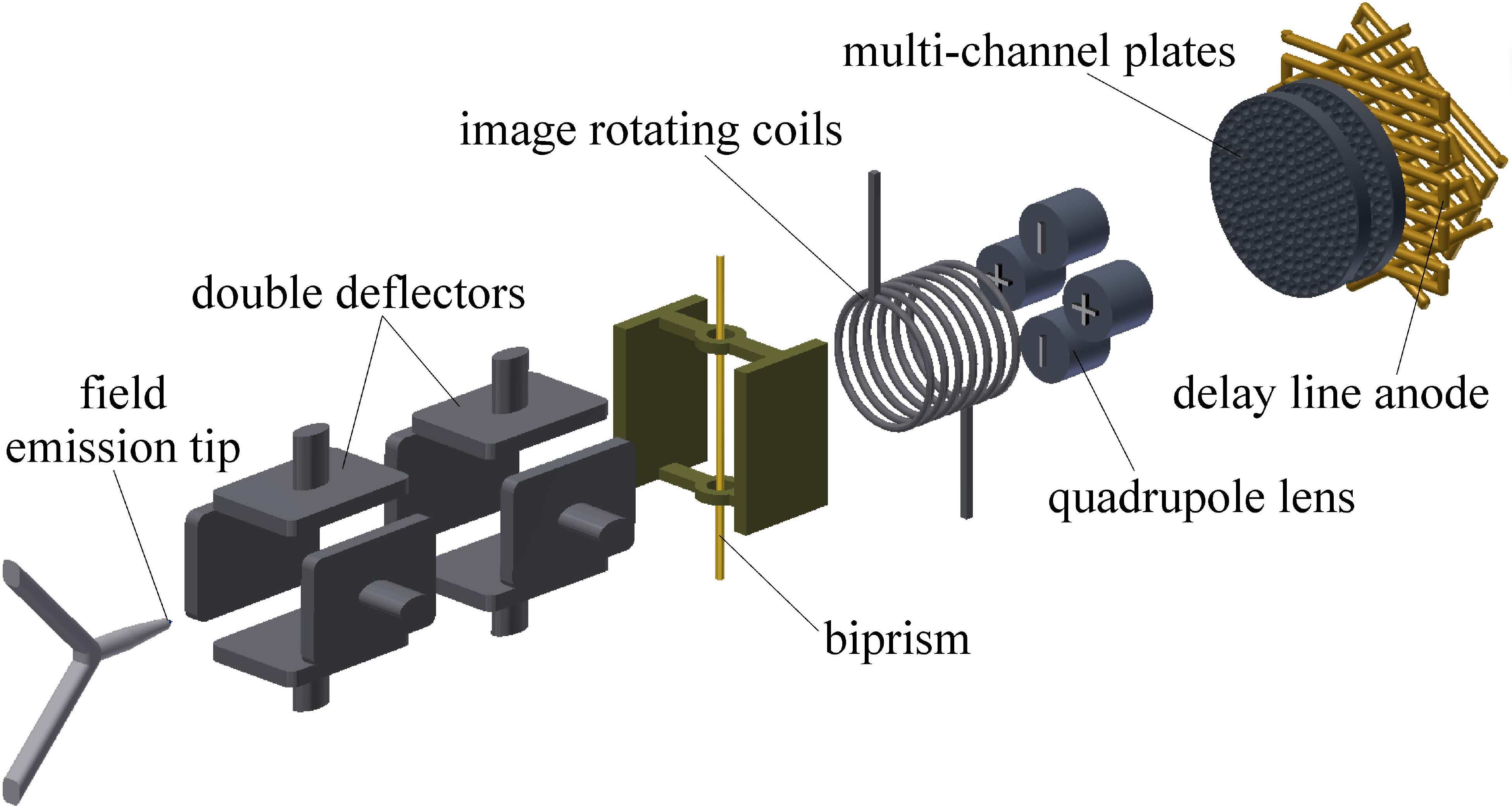} \caption{Sketch of the compact biprism electron interferometer (not to scale). The coherent electron beam is field emitted by a tungsten tip and guided by double deflectors. A biprism fiber separates and combines the partial matter waves that interfere at the entrance of a magnifying quadrupole. An image rotating coil can rotate the resulting pattern for alignment. The magnified interferogram is amplified by two multi-channel plates and detected by a delay line detector.
}
\label{fig1}
\end{figure}

\section{Experimental setup}\label{sec2}

The setup of the biprism interferometer is illustrated in fig.~\ref{fig1}. The source for coherent electrons is a field emission tip that can be prepared by pulsed etching of a polycrystalline tungsten wire \cite{Chang2012}. The beam gets aligned by two deflection electrodes to coherently illuminate a biprism fiber. It acts as a beamsplitter for the electron matter waves, if a small positive voltage is applied on the fiber. The electric force bends the separated beam paths towards each other and superimposes them in front of a magnifying quadrupole lens. The key feature of the electron biprism, analogue to the optical biprism, is that all possible beam paths get deflected by the same angle, leading to a common angle of superposition at the entrance of the quadrupole \cite{Mollenstedt1956a}. The superposition of the two partial beams leads to an interference pattern parallel to the biprism fiber. The interferogram has a typical pattern periodicity of several hundred nanometers and needs to be magnified by the quadrupole lens to fit the spatial resolution of the multi-channel plate (MCP) detector. Such lenses are optimal suited for biprism interferometry since it is only necessary to magnify the pattern in the direction normal to the interferences. A small misalignment of the biprism fiber towards the magnifying axis, can be corrected by an image rotating coil. In our experiment, a drift distance of \unit[169.8]{mm} between the exit aperture of the quadrupole lens and the MCPs of the detector was chosen. The electron signal gets amplified by the MCP and detected by a hexagonal delay line \cite{Roentdek} with a spatial and temporal resolution of $\sim$ \unit[100]{\textmu m} and $\sim$ \unit[0.1]{ns} \cite{Schoessler}.

\begin{figure}[t]
\centering
\includegraphics[width=0.5\textwidth]{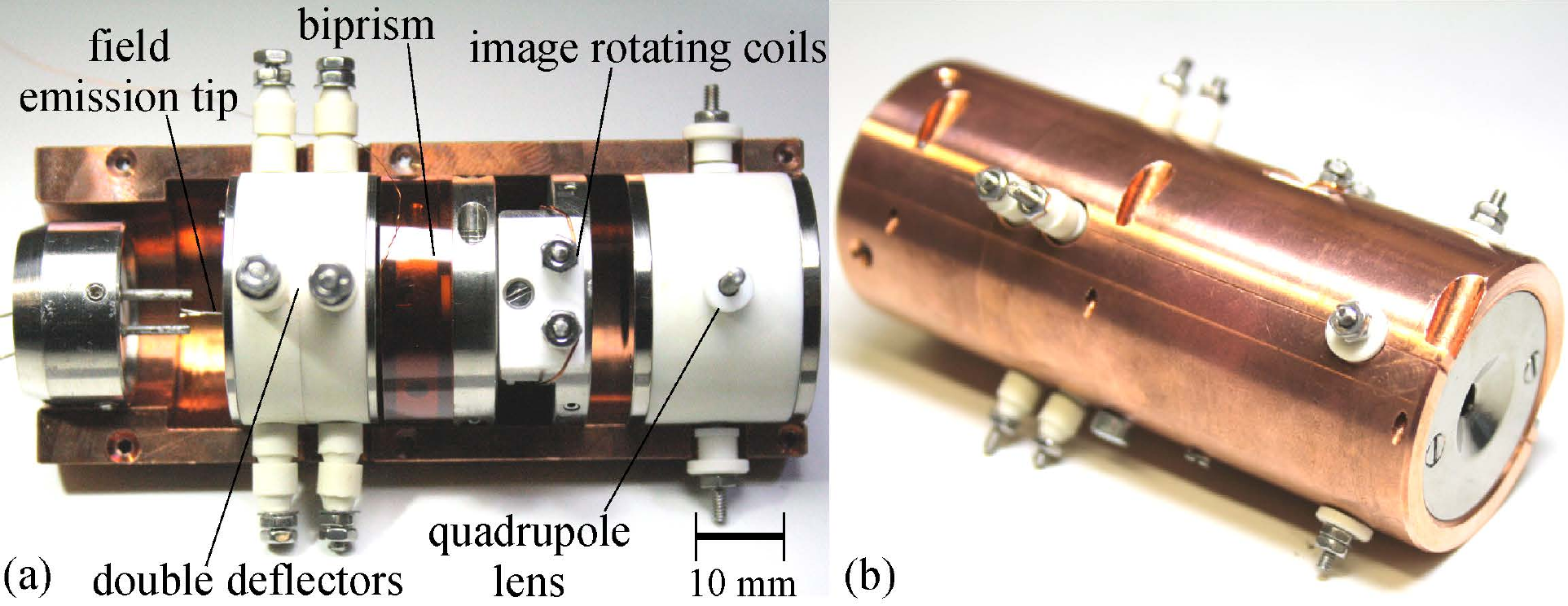} \caption{(a) Image of the components in the beam path from the tip to the quadrupole with the upper copper shell removed. (b) Interferometer parts from the tip to the quadrupole mounted together within both half-shells.}
\label{fig2}
\end{figure}

An image of the interferometer is shown in fig.~\ref{fig2} (a) and (b). The whole setup is fixed between two grounded copper half-shells with a length of \unit[87]{mm} and an outer diameter of \unit[36]{mm}. Some individual parts are identical or based on constructions used in interferometers by Hasselbach et al.~\cite{Hasselbach1988,Hasselbach2010}. The field emission tungsten tip is spot welded on a commercial holder and has a typical radius between \unit[10]{nm} and \unit[50]{nm}. The length of the tip profile and the radius can be controlled by the parameters in the etching process \cite{Chang2012}. The tip is positioned \unit[1]{mm} in front of an aperture of \unit[2]{mm} diameter acting as a grounded counter electrode. Its distance to the biprism is $a$ = \unit[19]{mm} and to the entrance of the quadrupole lens $b$ = \unit[28]{mm}. The tip emits electrons with an acceleration voltage of \mbox{$U_e$ = \unit[2250]{V}} into a double deflector module, consisting of four pairs of flat deflection electrodes with a length of \unit[5]{mm} and oriented around the beam with a distance of \unit[5.5]{mm} between them. The positive and negative voltage applied between two opposing electrode pairs is equal to keep a zero potential on the beam axis. The biprism fiber consists of a glass fiber with a diameter of \unit[400]{nm} that was manufactured by a special procedure described elsewhere \cite{Schuetz2014,Warken2007,Warken2008}. It is coated with a gold-palladium alloy to ensure a smooth, conductive surface and is glued on a holder isolated by a non-conductive foil. The fiber is positioned between two grounded titanium electrodes that are \unit[4]{mm} apart from each other. The rotating coil is winded around a \unit[7]{mm}-diameter tube. The quadrupole lens is made out of four opposing cylindrical electrodes with a length of \unit[10]{mm}, a diameter of \unit[7.6]{mm} and a distance normal to the beam path towards each other of \unit[6.7]{mm}. As for the double deflectors, the electrodes are mounted within isolating holders made out of \textsc{Macor}. An aperture with a diameter of \unit[4]{mm} is positioned at the entrance and exit of the quadrupole lens to decrease the amount of secondary stray electrons on the detector. The interferometer is magnetically shielded by a mu-metal tube and in a vacuum chamber at $\sim$\unit[5$\times$10$^{-10}$]{mbar} to ensure a long lifetime and stable emission of the tip \cite{Yeong2006}. 

\section{Experimental results}

\begin{figure}
	\centering
	\includegraphics[width=0.5\textwidth]{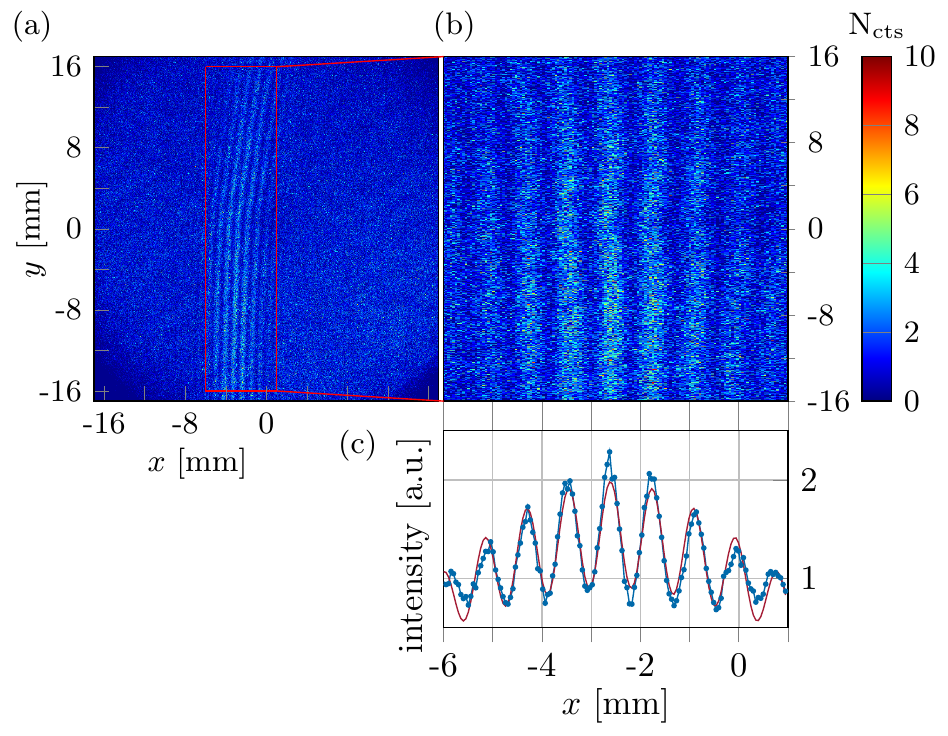} \caption{(a) Image of the interference pattern on the detector screen with the set experimental values $U_e$ = \unit[2250]{V}, $U_f$~=~\unit[(0.559+0.69)]{V} and quadrupole voltages \unit[+2960]{V} between the electrodes normal to the fringes and \unit[-2910]{V} parallel to them. (b) Focus of the region within the red rectangle where the fringes were straightened by a fourth degree polynomial. (c) Blue curve: average intensity along the $y$-direction on the screen. Red curve: numerical fit of the intensity.}
	\label{fig3}
\end{figure}

With this device it was possible to observe high contrast electron interference pattern such as shown in  fig.~\ref{fig3} (a) for 5$\times$10$^{5}$ detected particles. The temporal and spatial information of every detection event was recorded for a correlation analysis after data acquisition. The small bend in the middle of the pattern in fig.~\ref{fig3} (a) is due to a deviation of the beam from the optical axis in the quadrupole lens. It can be corrected by a fourth degree polynomial fit on the fringes and subsequent straightening. The resulting pattern in the marked red rectangle (\unit[7]{mm} $\times$ \unit[32]{mm}) is exhibited in fig.~\ref{fig3} (b). Thus, nine fringes are visible in the image. The color bar indicates the intensity distribution of the incoming particles. In fig.~\ref{fig3}~(c) the average intensity along the $y$-direction is plotted against the distance $x$ on the screen. The distribution is enfolded by a $sinc$-function due to the similarity to the double slit experiment analysis \cite{Hecht14}. Therefore, the data in fig.~\ref{fig3} (c) was fitted with the model function \mbox{$I(x)=I_0\cdot\left(1+K_m\cdot\cos(\frac{2\pi x}{s_m}+\phi_0)\right)\cdot {\rm sinc}^2(\frac{2\pi x}{s_1}+\phi_1)$} revealing a contrast $K_m$ = \unit[(37.2 $\pm$ 5)]{\%} and the fringe distance $s_m$ = \unit[(0.85 $\pm$ 0.02)]{mm}. The phases $\phi_0$, $\phi_1$, the average intensity $I_0$ and the width of the interference pattern $s_1$ are additional fitting parameters. The magnification of the interferogram in fig.~\ref{fig3} (a) is 2517 $\pm$ 6. It is given by the determined fringe distance $s_m$ on the detector after magnification divided by the theoretical fringe distance $s_0$ = \unit[338]{nm} at the entrance of the quadrupole, which is calculated with $s_0=\frac{\lambda_{dB}}{\theta}$, and $\theta=2 \gamma\cdot\frac{a}{a+b}$. Thereby, $\lambda_{dB}$ is the de Broglie wavelength of the electrons, $\theta$ the superposition angle at the entrance of the quadrupole, \mbox{$\gamma=\frac{\pi}{2\, ln (R/r)}\frac{U_f}{U_e}$} the deviation angle of the biprism fiber with an applied voltage of $U_f$, $r$ is the radius of the fiber and $R$ its distance to the grounded electrodes \cite{Lenz1984}.

We additionally performed beam path simulations with the program \textit{Simion} \cite{Simion}. The superposition angle was extracted from the simulation by a method described in \cite{Schuetz2015b}. It is known that contact potentials between the gold/palladium coating of the fiber and the titanium electrodes influence the effective potential interacting with the separated beams \cite{Krimmel1964,Bruenger1972}. For that reason, we adapted the biprism voltage until nine interference stripes fit in the superposition area. This is achieved by adding \unit[0.69]{V} in the simulation to the experimentally applied voltage of \unit[0.559]{V}. This extra voltage is considered to be the contact potential and agrees well comparing the literature values for the work functions of the averaged \unit[80:20]{\%} gold/palladium alloy and the one of titanium \cite{Walker}. Their difference amounts to \unit[0.71]{eV}. Our simulations revealed a superposition angle of \unit[7 $\times$ 10$^{-5}$]{rad} and a pattern periodicity of \unit[369]{nm} before magnification. The quadrupole magnification was simulated to be 2886, which can be considered as an upper bound for perfect beam alignment on the optical axis. The reasonable small variations to the theoretical fringe distance $s_0$ and magnification are possibly due to the neglected beam adjustment voltages and small deviations between the simulated and experimental setup distances.

\begin{figure}[t]
	\centering
	\includegraphics[width=0.5\textwidth]{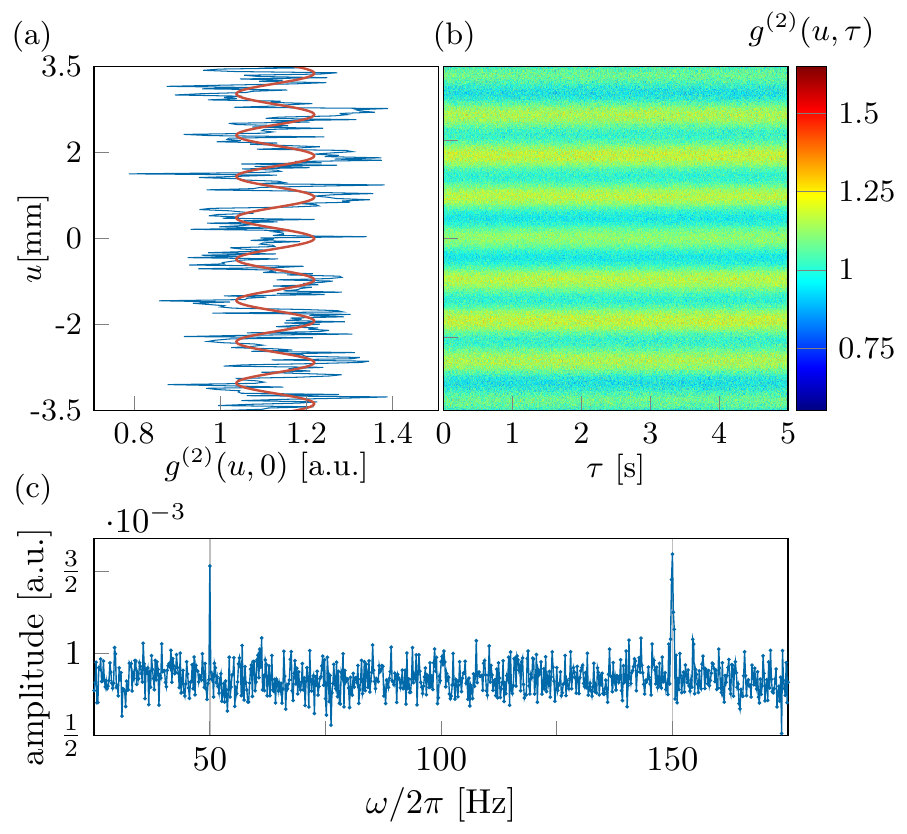} \caption{(a) Blue line: the second-order correlation function $g^{(2)}(u,0)$ at the correlation time $\tau=0$. Red curve: fit function to reveal the contrast and the pattern periodicity. (b) The second-order correlation function $g^{(2)}(u,\tau)$ for $\tau$ ranging between 0 and 5 seconds. $g^{(2)}(u,\tau)$ is extracted from the data in fig.~\ref{fig3} (b). The periodic fringe pattern is clearly observable. (c) Plot of the amplitude spectrum calculated via a numerical Fourier transformation of $g^{(2)}(u=M_u\cdot s_g/2,\tau), M_u\in\mathrm{N_0}$ and subsequent averaging. Periodic perturbations of the interference fringes in time become visible and identify two characteristic frequencies at \unit[50]{Hz} and \unit[150]{Hz}.}
	\label{fig4}
\end{figure}
As recently demonstrated \cite{Guenther2015,Rembold2017} and discussed above, contrast reducing dephasing effects, such as electromagnetic oscillations or mechanical vibrations can be isolated and corrected by a second-order correlation analysis of the measured interference pattern. We applied this method, described in detail elsewhere \cite{Guenther2015,Rembold2017,Rembold2017b}, to the interference data in fig.~\ref{fig3} (b). The analysis provides the second-order correlation function $g^{(2)}(u,\tau)$ with the correlation length $u$ and correlation time $\tau$ between the detected particles. Consequentially, we can determine the fringe distance $s_g$ and the contrast $K_g$ of the unperturbed interference pattern at the temporal position $\tau=0$, $g^{(2)}(u,0)$, as shown in fig.~\ref{fig4} (a). The red curve is a fit function $g^{(2)}(u,0)=1+\frac{K_g^2}{2}\cdot \cos{(\frac{2\pi u}{s_g}+\phi_g)}+O$ with the fit parameters contrast $K_g$, the fringe distance $s_g$, the phase $\phi_g$ and the offset $O$ \cite{Rembold2017b}. We obtain $s_g$ = \unit[(0.84 $\pm$ 0.02 )]{mm} and $K_g$ = \unit[(42.5 $\pm$ 7)]{\%}. The contrast $K_g$ is higher than $K_m$ suggesting that disturbing effects may wash-out the interference pattern. The significant high standard deviations for both values are due to a limited number of detected particles and therefore a large noise. In fig.~\ref{fig4}~(b) $g^{(2)}(u,\tau)$ is shown for correlation times up to five seconds. Along the $u$-axis the unperturbed fringe pattern with a distance $s_g$ can be observed. The amplitude spectrum of the correlation function, $|\mathcal{F}(g^{(2)}(u,\tau))(u,\omega)|$, reveals the perturbation characteristics. It is determined using a numerical temporal Fourier transformation at the spatial positions $(u=M_u\cdot s_g/2,\tau, M_u\in\mathbb{N}_0)$, where the correlation function has its maximum signal \cite{Rembold2017, Rembold2017b}. In fig.~\ref{fig4}~(c) the average over all amplitude spectra calculated at the spatial positions $u=M_u\cdot s_g/2$ is plotted. For optimal settings of the spatial and temporal discretization step size of the numerical correlation function \cite{Rembold2017b}, two clear peaks at \unit[50]{Hz} and \unit[150]{Hz} evolve. These frequencies affect the electron waves and lead to unwanted contrast loss. Their spatial perturbation amplitude before the magnification can be determined according to the description in \cite{Rembold2017} to \unit[14.3]{nm} at \unit[50]{Hz} and \unit[14.8]{nm} at \unit[150]{Hz}. Thereby, it is assumed that the perturbation takes place before magnification, which is reasonable since the amplitudes are low. They probably originate from the utility frequency of the electrical power supplies. Their detection demonstrates the usability of our device as a sensor for external perturbation frequencies. 

\begin{figure}[t]
	\centering
	\includegraphics[width=0.49\textwidth]{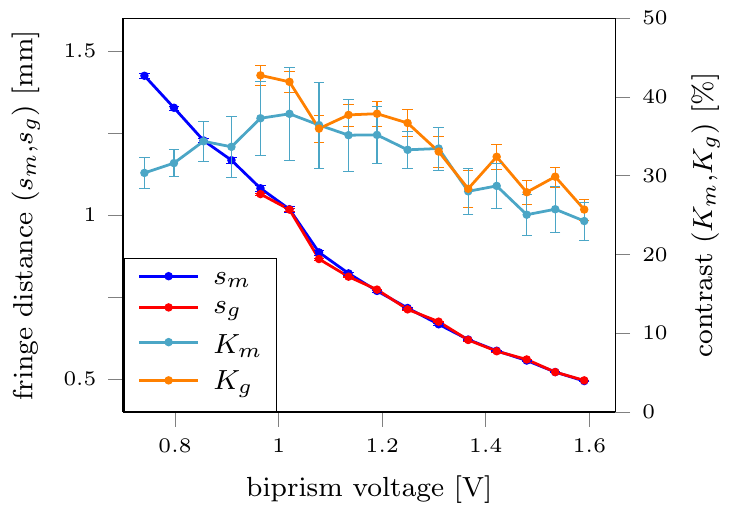} \caption{Fringe distances $s_m$ and $s_g$ together with the interference contrasts $K_m$ and $K_g$ as a function of the biprism voltage. The data is evaluated by spatial integration and correlation analysis. The error bars for $s_m$ and $s_g$ are smaller than the dot size. $K_m$ is in most cases lower than $K_g$ due to the influence of the detected dephasing perturbations.}
	\label{fig5}
\end{figure}

For further characterization of the compact interferometer, we present a series of measurements with variable voltages $U_f$ at the biprism fiber, going from \unit[(0.051+0.69)]{V} to \unit[(0.900+0.69)]{V} in $\sim$ \unit[50]{mV} steps. With those sixteen data files, the same analysis is executed as described above. The data is shown in fig.~\ref{fig5}, whereas $U_f$ is plotted against both the fringe distance and the contrast. The values for $s_g$ and $K_g$ refer to the unperturbed case originating from the second-order correlation analysis. As expected from theory \cite{Lenz1984}, the fringe distance decreases with increasing biprism voltage, since the superposition angle of the partial waves increases. This is observed for both $s_m$ and $s_g$ leading to nearly the same results. The contrast $K_m$ increases up to $\sim$~\unit[38]{\%} at $\sim$ \unit[1]{V} biprism voltage. For lower voltages $K_m$ decreases due to diffraction effects at the edge of the fiber. For higher voltages a reduced contrast is expected due to coherence considerations with a  beam source of finite extent \cite{Lenz1984,Maier1997}. $K_g$ reveals a similar curve progression with a maximum of \unit[42.7]{\%}. As expected, the contrast $K_g$ is at most voltages higher than the contrast $K_m$ disturbed by the two dephasing frequencies. The first four measurements of $K_g$ can not be evaluated with the $g^{(2)}$ analysis because of the influence of diffraction. It is not included in the theory of the $g^{(2)}$ analysis and significant for small biprism voltages.

\section{Conclusion}

We demonstrated a compact biprism matter wave interferometer for free electrons providing interference fringes with a contrast up to \unit[42.7]{\%}. The dimension and the robustness of the setup is sufficient to be integrated in a mobile device. The interference data was recorded by a delay line detector with a high spatial and temporal single-particle resolution. This allowed a second-order correlation and Fourier analysis revealing the undisturbed contrast and the pattern periodicity of the interferogram. It was compared to the values obtained by pure spatial signal integration. Our device allowed the identification of dephasing frequencies at \unit[50]{Hz} and \unit[150]{Hz} and therefore demonstrated its applicability for the detection of external perturbations with the recently developed second-order correlation data analysis \cite{Rembold2017,Guenther2015,Rembold2014,Rembold2017b}. Identifying dephasing oscillations from the lab environment is a very helpful tool to improve experiments for sensitive phase measurements. The setup can also be applied to test the coherence of novel beam sources \cite{Ehberger2015,Hommelhoff2006a}. Especially the simple design and the small dimensions make the interferometer easy to handle and usable in various environments. The sensitivity towards vibrational or electromagnetic dephasing and inertial forces such as rotation and acceleration could be increased significantly with a larger beam path separation. This would require an interferometer scheme with two or three biprism fibers in combination with a quadrupole or an einzel-lens \cite{Schuetz2015b,Clauser1988,Hasselbach1993}. 

\section{Acknowledgements}

This work was supported by the Vector Stiftung and the Deutsche Forschungsgemeinschaft through the Emmy Noether program STI 615/1-1 and the research grant STI 615/3-1. A.R. acknowledges support from the Evangelisches Studienwerk e.V. Villigst. The authors thank N.~Kerker, R.~R\"{o}pke and G.~Sch\"{u}tz for helpful discussions.

\end{document}